\documentclass[fleqn,usenatbib]{mnras}

\usepackage{newtxtext,newtxmath}

\usepackage[T1]{fontenc}

\DeclareRobustCommand{\VAN}[3]{#2}
\let\VANthebibliography\thebibliography
\def\thebibliography{\DeclareRobustCommand{\VAN}[3]{##3}\VANthebibliography}

\usepackage{graphicx}	
\usepackage{amsmath}	

\title[The X-ray transient KY TrA]{Evidence for a black hole in the historical X-ray transient A 1524-61 (=KY TrA)}

\author[I. V. Yanes-Rizo et al.]{
I. V. Yanes-Rizo$^{1,2}$\thanks{E-mail: idairayr@iac.es},
M. A. P. Torres$^{1,2}$,
J. Casares$^{1,2}$,
M. Monelli$^{1,2}$,
P. G. Jonker$^{3,4}$,
T. Abbot$^{5}$, 
\newauthor  M. Armas Padilla$^{1,2}$,
T. Muñoz-Darias$^{1,2}$
\\
$^{1}$Instituto de Astrofísica de Canarias, E-38205 La Laguna, S/C de Tenerife, Spain\\
$^{2}$Departamento de Astrofísica, Universidad de La Laguna, E-38206 La Laguna, S/C de Tenerife, Spain\\
$^{3}$Department of Astrophysics\,/\,IMAPP, Radboud University, Heyendaalseweg 135, NL-6525 AJ Nijmegen\\
$^{4}$SRON, Netherlands Institute for Space Research, Niels Bohrweg 4, 2333 CA, Leiden, The Netherlands\\
$^{5}$AURA\,/\,CTIO, Casilla 603, La Serena, Chile\\
}
\date{Accepted XXX. Received YYY; in original form ZZZ}

\pubyear{2023}

\begin{document}
\label{firstpage}
\pagerange{\pageref{firstpage}--\pageref{lastpage}}
\maketitle

\begin{abstract}
We present VLT spectroscopy, high-resolution imaging and time-resolved photometry of KY TrA, the optical counterpart to the X-ray binary A 1524-61. We perform a refined astrometry of the field, yielding improved coordinates for KY TrA and the field star interloper of similar optical brightness that we locate $0.64 \pm 0.04$\,arcsec SE. From the spectroscopy, we refine the radial velocity semi-amplitude of the donor star to $K_2 = 501 \pm 52$\,km s$^{-1}$ by employing the correlation between this parameter and the full-width at half-maximum of the H$\alpha$ emission line. The $r$-band light curve shows an ellipsoidal-like modulation with a likely orbital period of $0.26 \pm 0.01$\,d ($6.24 \pm 0.24$\,h). These numbers imply a mass function $f(M_1) = 3.2 \pm 1.0$\,M$_\odot$. The KY TrA de-reddened quiescent colour $(r-i)_0 = 0.27 \pm 0.08$ is consistent with a donor star of spectral type K2 or later, in case of significant accretion disc light contribution to the optical continuum. The colour allows us to place a very conservative upper limit on the companion star mass, $M_2 \leq 0.94$\,M$_\odot$, and, in turn, on the binary mass ratio, $q = M_2/M_1 \leq 0.31$. By exploiting the correlation between the binary inclination and the depth of the H$\alpha$ line trough, we establish $i = 57 \pm 13$\,deg. All these values lead to a compact object and donor mass of $M_1 = 5.8^{+3.0}_{-2.4}$$\,M_\odot$ and $M_2 = 0.5 \pm 0.3$\,M$_\odot$, respectively, thus confirming the black hole nature of the accreting object. In addition, we estimate a distance toward the system of $8.0 \pm 0.9$\,kpc.

\end{abstract}




\section{Introduction}

Low-mass X-ray binaries (LMXBs) are binary systems containing a black hole or a neutron star accreting matter from a low-mass Roche-lobe filling companion star via an accretion disc. A subgroup of LMXBs, dubbed X-ray transients (XRTs), are known for showing unpredictable episodes of increased bolometric luminosity caused by enhanced mass accretion onto the compact object \citep[see e.g.][]{mcclintock2006}. These outburst episodes are followed by long periods of low accretion luminosity called quiescent state when the companion star may be detectable over the diminished disc emission. This provides an opportunity to perform dynamical mass measurements by establishing the orbital period $P$, the radial velocity semi-amplitude of the companion star $K_2$, the binary inclination $i$ and the mass ratio $q$ \citep[see][for a review]{casaresjonker2014}.

When the companion is not detected, alternative techniques have to be considered to derive reliable mass constraints. This is the case of the XRT A 1524-61, subject of this work, which lacks a dynamical study due to the faintness of the quiescent optical counterpart and the presence of an interloper star. 

A 1524-61 was discovered on 12 November 1974 with the \emph{Ariel V} satellite \citep{pounds1974}. Based on its soft X-ray spectral properties and optical/X-ray similarities with A0620-00, this new XRT was proposed to host a black hole accretor \citep{white1984, murdin1977}. Its optical counterpart (KY TrA) was identified during the discovery outburst with a $B = 17.5$ star \citep{murdin1977}. Further support for the presence of a black hole in KY TrA came from X-ray spectral properties observed during a new low-intensity outburst detected by the SIGMA telescope in Aug 1990 \citep{barret1992}. After decades of oblivion, KY TrA was again recovered in 2004-2010 during quiescence \citep{zurita2015}. These authors showed through inspection of H$\alpha$ and $I$-band frames that KY TrA is the NW component of a pair of stars separated $\sim 0.4$\,arcsec \footnote{\citet{zurita2015} erroneously reported in their abstract an offset of $\sim 1.4$\,arcsec. The correct units are pixels with $1$\,pix $= 0.252$\,arcsec.} having a brightness of $R = 22.3$\,, $I = 21.5$ (corrected for light from the interloper star). From photometric constraints on the companion spectral type, \citet{zurita2015} suggested an orbital period of $ \approx$$8$\,h, with a robust upper limit at $15$\,h. In addition, they derived $K_2$ to be $630 \pm 74$\,km s$^{-1}$ by using the H$\alpha$ emission line detected in a single poor quality spectrum obtained during quiescence. These numbers imply a tentative mass function in the range of $\sim 9 - 16$\,M$_\odot$. 

In this paper we present $0.5$\,arcsec spatial resolution imaging, time-series photometry and higher quality spectroscopy of the optical quiescent counterpart to A 1524-61. We improve the astrometric locations of KY TrA and its interloper, identify the likely orbital period and resolve the double-peaked morphology of the H$\alpha$ emission line from the quiescent accretion disc. We exploit the properties of the H$\alpha$ emission line to constrain the orbital parameters. In what follows we detail the observations and data reduction steps (Section~\ref{sec:2}). The data analysis and results are given and discussed in Section~\ref{sec:3}. Finally, in Section~\ref{sec:4} we constrain the orbital inclination and discuss the implications on the compact object mass.

\section{Observations and data reduction}
\label{sec:2}
\subsection{Spectroscopy}
\label{sec:spectroscopy} 
Spectroscopy for KY TrA was acquired on 2016 April 4 and 7 in Service mode at ESO's Paranal observatory (Chile). The observations were performed with the FOcal Reducer and low dispersion Spectrograph 2 \citep[FORS2,][]{appenzeller1998} which was attached to the Cassegrain focus of the 8.2-m Unit 1 Very Large Telescope (VLT). Three consecutive $820$\,s\ spectra were collected each night with the grism GRIS 600RI and a $0.7$\,arcsec slit. In a best effort to have both KY TrA and the interloper on the slit, its position angle PA was set to $-15$ and $-20$\,deg during the first and second night, respectively. The instrument was used with the standard resolution collimator and the two $2048\times4096$ pixels MIT CCDs binned by a factor of two. This instrumental setup provided a $5015-8330$\,\AA\ spectral coverage with a 1.6\,\AA\ pixel$^{-1}$ dispersion and a nominal slit width limited resolution of $\sim$ 3.3\,\AA\ full-width at half-maximum (FWHM). The seeing, measured from the spectral trace of a field star centered on the slit, ranged between $0.5-0.7$\,arcsec during the observations. 

The reduction of the data was performed with standard routines implemented in {\sc pyraf}\footnote{\url{https://github.com/iraf-community/pyraf}}. It consisted of de-biasing and flat-fielding the science and calibration arc-lamp frames. The latter were obtained at the end of each observing night in order to perform the wavelength calibration of the data. The pixel-to-wavelength scale was derived through two-piece cubic spline fit to 17 arc lines. The root mean square (rms) scatter of the fit was $< 0.04$\,\AA. Wavelength zeropoint deviations were corrected by applying to the spectra zeropoint shifts calculated using the [O\,{\sc i}] 6300.3\,\AA\ sky line. Given the light contamination of KY TrA by the interloper, the spectral profiles were not traced to avoid potential large departures from the target location during the extraction. Instead, the aperture was fixed at the position of the target spatial profile at H$\alpha$ and sized for each spectrum to maximize the number of pixels containing signal from the object while minimizing the interloper contamination. The extracted individual spectra have a mean continuum signal-to-noise ratio (SNR) of 3 in the $5800-6300$\,\AA\ wavelength region, reaching $\sim 8$ at maximum intensity of the H$\alpha$ emission line. The data were imported to the {\sc molly}\footnote{\textsc{molly} was written by T.~R. Marsh and is available from \url{https://cygnus.astro.warwick.ac.uk/phsaap/software/molly/html/INDEX.html}.} package where we shifted them to the heliocentric rest frame and performed all subsequent analysis. For this, the spectra were normalised to the continuum level with three-knot spline fits after masking out the telluric bands and emission lines.

\subsection{Photometry}
\label{sec:photometry}
\label{sec:2.2}
\begin{figure}
	\includegraphics[width=\columnwidth]{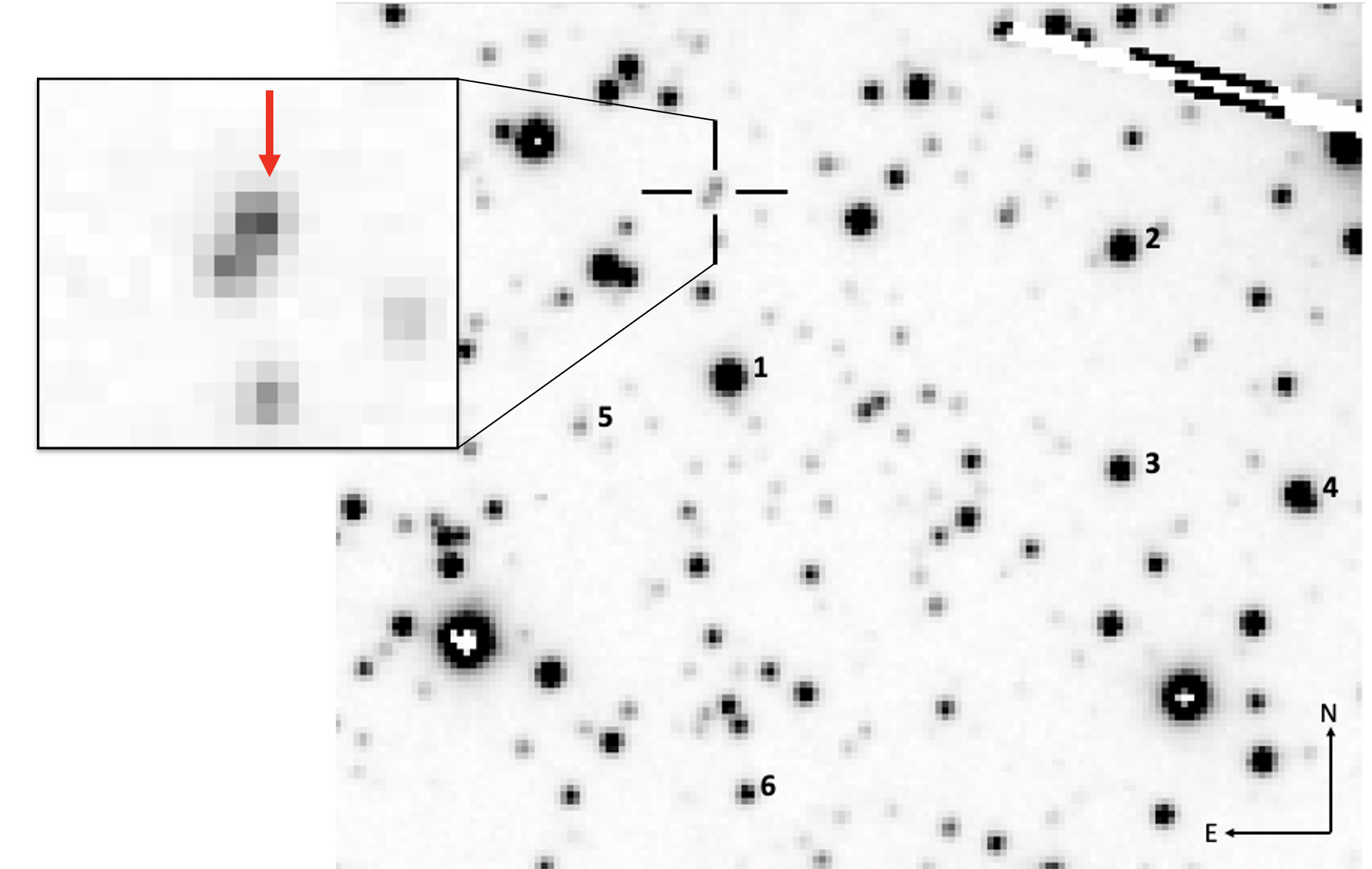}
    \caption{FORS2 120\,sec $I$-band acquisition image of the field containing KY TrA. The field of view is $40 \times 34$\,arcsec and the plate scale $0.25$\,arcsec\ pixel$^{-1}$. The position of KY TrA and its interloper are indicated by a black cross while the field stars used for the photometry are numbered following Table \ref{tab:table}. The target location has been zoomed in to show the clear distinction between KY TrA (NW source marked with a red arrow) and its interloper (SE).}
    \label{fig:finder}
\end{figure}

We also performed time-resolved photometry of KY TrA on the nights of 2019 May 10 and 11, using the Dark Energy Camera \citep[DECam;][]{flaugher2015} mounted on the prime focus of the Víctor M. Blanco 4-meter Telescope at Cerro Tololo Inter-American Observatory. DECam has 62 $2048\times 4096$ pixel CCD chips covering a field of view of $3$\,deg$^2$ with  a $0.26$\,arcsec\ pixel$^{-1}$ plate scale. KY TrA was placed on the central chip \#28. A total of 73 and five 200 s images were obtained in the $r$- and $i$-bands, respectively, covering over $\sim 9$\,h per night. The seeing ranged between 0.8 and 1.8 arcsec. The bias subtraction and the flat field correction of the CCD containing KY TrA were made using {\sc pyraf}. Photometry was carried out with the HiPERCAM\footnote{\url{https://github.com/HiPERCAM/hipercam}} reduction pipeline. Apertures were centered on the unresolved transient and interloper, and six field stars (Fig.~\ref{fig:finder}). After several tests, the aperture radius was set to six pixels ($1.6$\,arcsec) to sum the flux from KY TrA+interloper and maximize its signal-to-noise ratio. Seven $r$-band images with seeing larger than $1.5$\,arcsec provided only photometric upper limits and were rejected. Our photometry was calibrated using the Dark Energy Camera Plane Survey 2 \citep[DECaPS2; ][]{saydjari2023}. For an independent test we also used the ATLAS All-Sky Reference Catalog \citep{tonry2018} and obtained consistent results. Table \ref{tab:table} shows the $r$ magnitudes for the comparison star used in the differential photometry (star 1) and the five field stars used as reference for testing the stability of the light curves (stars 2-6).

\begin{table}
	\centering
	\caption{Mean $r$-band magnitudes and rms scatter for the six comparison stars obtained from the DECam light curves.}
	\label{tab:table}
	\begin{tabular}{ | c | c | } 
		\hline
            Star & $r$ \\ \hline 
		$1$ & 
	    \begin{tabular}{c} $18.39 \pm 0.01$ \\
            \end{tabular} \\ \hline
		$2$ &
		\begin{tabular}{c} $19.15 \pm 0.01$ \\
            \end{tabular} \\ \hline
            $3$ &
		\begin{tabular}{c} $19.88 \pm 0.01$ \\
            \end{tabular} \\ \hline
            $4$ &
		\begin{tabular}{c} $19.51 \pm 0.01$ \\
            \end{tabular} \\ \hline
            $5$ &
		\begin{tabular}{c} $22.47 \pm 0.08$ \\
            \end{tabular} \\ \hline
            $6$ &
		\begin{tabular}{c} $21.93 \pm 0.05$ \\
            \end{tabular} \\ \hline
	\end{tabular}
\end{table}

\section{Data analysis and results}
\label{sec:3}

\subsection{Improved astrometric solution}
KY TrA and its interloper are clearly distinguished in a 120\,sec $I$-band FORS2 acquisition image taken on 2016 April 07 during the spectroscopic campaign. The image of KY TrA was recorded on CCD1 under 0.5\,arcsec seeing, it has a field of view of $7 \times 4$\,arcmin${^2}$ and it is sampled with a $0.252$\,arcsec\ pixel$^{-1}$ plate scale (see Fig.~\ref{fig:finder}). We performed astrometry on this image using Gaia Data Release-2 stars \citep{gaia2018} employing the Gaia\footnote{\url{https://github.com/Starlink/starlink/tree/master/applications/gaia}} image tool to fit the positions of 102 Gaia point sources that are not saturated ($g > 17$) in our image and that delivered an astrometric solution with rms $0.043$\,arcsec. We determined the coordinates of KY TrA to be $\alpha$\,(J2000) $= 15{:}28{:}16.93$ and $\delta$\,(J2000)$= -61{:}52{:}57.95$.
The coordinates of the interloper are $\alpha$\,(J2000) $= 15{:}28{:}16.97$ and $\delta$\,(J2000)$= -61{:}52{:}58.52$. The separation between the two components is $0.64 \pm 0.04$\,arcsec. These determinations are based on the positions derived from point spread-function fitting (see below). They supersede previous values inferred from the centroid of the KY TrA+interloper pair measured in H$\alpha$ and $I$-band images of lower quality \citep{zurita2015}. As we are able to resolve the blend we performed Point-Spread Function (PSF) photometry in order to compare with the $I$-band magnitude of KY TrA in \citet{zurita2015}. The DAOPHOT package \citep{stetson1987} was used with a Moffat distribution model for eleven stars after removing their neighbours. Differential photometry was performed relative to stars in the DECaPS2 catalogue, which $I$-band magnitudes were calculated using the transformations reported by Lupton (2005)\footnote{\url{http://classic.sdss.org/dr4/algorithms/sdssUBVRITransform.html}}. We obtained $I = 21.3 \pm 0.3$, where the uncertainty is dominated by the photometric errors. This is consistent with the value reported by \citet[][$I = 21.47 \pm 0.09$]{zurita2015}.

\subsection{The H$\alpha$ emission line and inferred $K_2$ and $q$}
\label{sec:3.1}
\begin{figure}
	\includegraphics[width=\columnwidth]{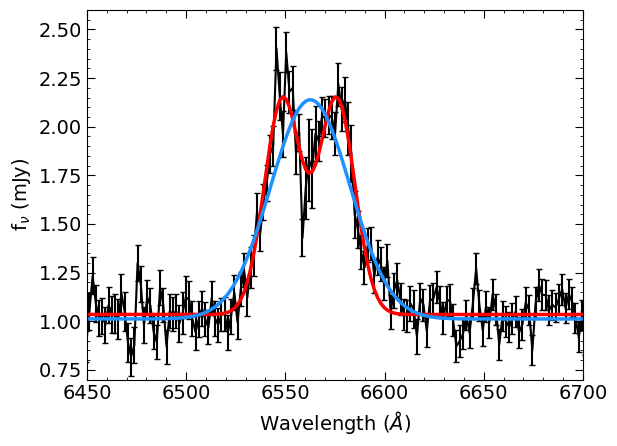}
    \caption{The average H$\alpha$ emission line profile at the heliocentric rest frame is shown in black \textbf{with associated uncertainties overplotted}. Blue and red lines represent the best-fit single- and double-Gaussian models, respectively.}
    \label{fig:halfa_fit}
\end{figure}

H$\alpha$ emission is the only discerned spectral feature from KY TrA in the FORS2 data. For the subsequent analysis, the normalized individual spectra were averaged using inverse variance weights to maximize the SNR of the resulting sum, which is shown in Fig.~\ref{fig:halfa_fit}. Contrary to \citet{zurita2015}, the double-peaked morphology of the emission line is clearly resolved. Thus, we can infer the fundamental parameters $K_2$ and $q=M_2/M_1$, where $M_1$ and $M_2$ stand for the masses of the compact star and its companion, respectively. To do this we exploit the H$\alpha$ correlations found in \citet{casares2015, Casares2016}. Following these works, we fitted both a single and a symmetric double-Gaussian model to the individual and averaged spectra contained within $\pm 10000$\,km s$^{-1}$ of the H$\alpha$ line rest wavelength. Before conducting the fits, these models were degraded to the $3.3$\,\AA\ instrumental resolution of our average spectrum. The fits to the averaged profile are illustrated in Fig.~\ref{fig:halfa_fit}. 

We start by employing the correlation that links the radial velocity semi-amplitude of the donor star with the FWHM of the H$\alpha$ emission line in quiescent XRTs: $K_2 = 0.233(13) \times {\rm FWHM}$ \citep{casares2015}. From the single Gaussian fits to the individual profiles we obtain ${\rm FWHM} = 2114 \pm 157$\,km s$^{-1}$, where the value and the uncertainty correspond to the mean and the standard deviation, respectively. Alternatively, a fit to the average profile yields ${\rm FWHM} = 2151 \pm 62$\,km s$^{-1}$. Given that the spectroscopy only covers a small fraction of the orbit, we decided to adopt the fit to the average profile but adding quadratically a 10 per cent uncertainty to account for the intrinsic FWHM variability that is typically observed in XRTs \citep[see][]{casares2015}. This leads to ${\rm FWHM} = 2151 \pm 224$\,km s$^{-1}$ and, thus, the FWHM-$K_2$ correlation yields $K_2 = 501 \pm 52$\,km s$^{-1}$. This is consistent within $2 \sigma$ with the value $630 \pm 74$\,km s$^{-1}$ reported in \citet{zurita2015}, which was obtained through a single, lower-quality spectrum. 

Next, we constrain the mass ratio $q$ by making use of the existing correlation between this parameter and the ratio of the double-peak separation (DP) to FWHM of the H$\alpha$ line. This relation is given by $\log q = -6.88(0.52) - 23.2(2.0)$ $\log\left(\frac{\rm DP}{\rm FWHM}\right)$ \citep{Casares2016}. From the fit of the single and double Gaussian to the average spectrum we obtain ${\rm DP} = 1247 \pm 23$\,km s$^{-1}$ and ${\rm FWHM} = 2151 \pm 62$\,km s$^{-1}$. The calculation of $q$ was done through Monte Carlo randomization, where ${\rm DP}/{\rm FWHM}$ is treated as being normally distributed around their measured value with standard deviation equal to its uncertainty. Thus, we obtain $q = 0.04_{-0.03}^{+0.05}$. Nevertheless, given the limited phase coverage, this value could be biased as it is not accounting for possible variability of the emission line throughout the orbit. In Section \ref{sec:spectraltype}, we provide a more conservative constraint on $q$ by restricting the spectral type of the donor star through colour information.

\subsection{Optical light curve and search for periodicities}
\begin{figure}
	\includegraphics[width=\columnwidth]{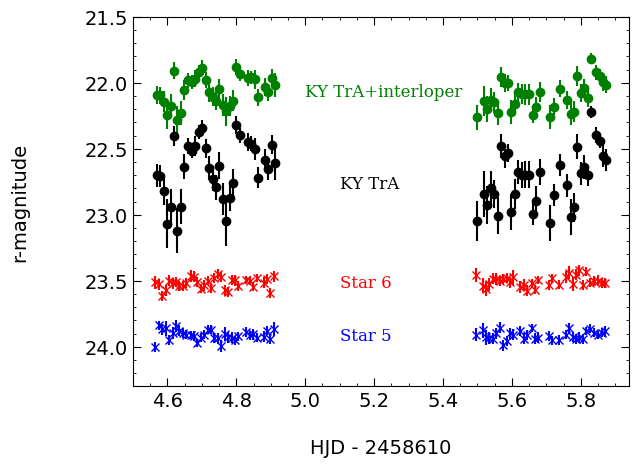}
    \caption{$r$-band light curve of the unresolved KY TrA+interloper pair (green points), KY TrA after correcting for the interloper light contamination (black points) as well as the field stars 6 (red points) and 5 (blue points) which have similar brightness to the blended pair of stars and KY TrA, respectively. These has been shifted $1.6$ and $1.5$\,mag for display purposes. The photometric uncertainties in KY TrA include the error contribution from the interloper flux removal and photometric system conversion.}
    \label{fig:LC}
\end{figure}

The DECam $r$-band light curve extracted from our aperture photometry (Section \ref{sec:2.2}) contains the light of both KY TrA and the unresolved interloper (see Fig.~\ref{fig:LC}). From the time-series photometry we measure mean magnitudes and rms variability of $r = 22.07 \pm 0.11$ and $i = 21.45 \pm 0.07$ for the blended light. The field star 6 has similar brightness ($r = 21.94 \pm 0.05$) and its $r$-band light curve is also plotted in Fig.~\ref{fig:LC} for comparison. The amplitude of the variability intrinsic to KY TrA is veiled by the presence of the interloper that according to the photometry presented in \citet{zurita2015} has $I = 21.78 \pm 0.09, R = 22.8 \pm 0.1$ and $V = 23.6 \pm 0.1$. We converted all these Johnson-Cousins magnitudes to the SDSS photometric system (Lupton 2005) and use them to remove the contaminating flux from the interloper. After this correction we establish mean magnitudes and rms variability for KY TrA of $r = 22.7 \pm 0.2$ and $i = 22.0 \pm 0.1$. Our mean values are fully consistent with the magnitudes in \citet{zurita2015} ($r = 22.8 \pm 0.2$\,, $ i = 22.0 \pm 0.1$, where the uncertainties include statistical and photometric conversion errors). This confirms the quiescent state of KY TrA at the time of the DECam observations. Fig.~\ref{fig:LC} shows the contamination-corrected $r$-band light curve of KY TrA. For comparison purposes, we also plot the light curve of the field star 5 with close brightness and colour ($r = 22.47 \pm 0.08$\,, $ i = 21.74 \pm 0.07$) to KY TrA. A clear intrinsic modulation is seen for KY TrA in the first night, showing two maxima and two minima. The first minimum seems to be slightly fainter than the second, a telltale sign of an ellipsoidal modulation. The modulation is less clear in the second night, perhaps due to a larger level of flickering activity \citep{zurita2003}.

\begin{figure}
	\includegraphics[width=\columnwidth]{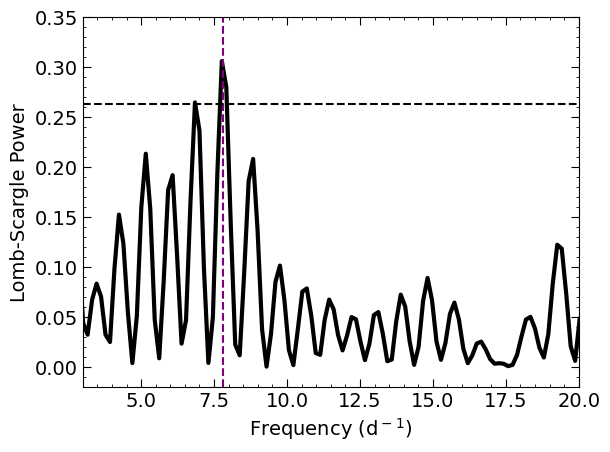}
    \caption{Lomb-Scargle periodogram from the $r$-band light curve. The 95 per cent white noise significance level has been represented by an horizontal black dashed line. The vertical dashed line marks the frequency at 7.8 cycle d$^{-1}$ (0.128-d period).}
    \label{fig:periodogram}
\end{figure}

\begin{figure}
	\includegraphics[width=\columnwidth]{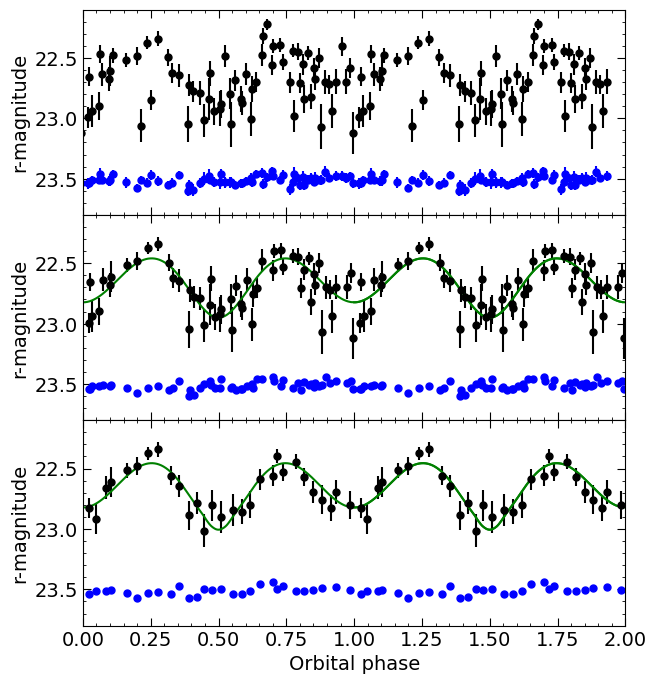}
        \caption{Top: The $r$-band light curves of KY TrA (black points) and star 5 (blue points) folded on the $0.26$\,d periodicity. The latter has been shifted $1.0$\,mag for display purposes. Middle: phase folded light curve after performing the $2 \sigma$ clipping procedure described in Section \ref{sec:4.2}. Bottom: the previous light curve co-added into 33 phase bins. Best-fitting synthetic models (solid green line) are included to guide the reader's eye for the global trend. One orbital cycle is repeated for clarity.}
    \label{fig:phasefolded}
\end{figure}

In order to identify the orbital period of KY TrA, we computed a Lomb-Scargle periodogram \citep{Lomb1976, Scargle1982} of the $r$-band light curve (see Fig.~\ref{fig:periodogram}). The frequency of the highest peak is found at 7.8 cycle d$^{-1}$ (0.128-d period), being above the 95 per cent white noise significance level. The second highest peak, at 6.9 cycle d$^{-1}$ (0.145-d period), is slightly above that level. To check the robustness of this period measurement, a bootstrap test has been computed with 1000 periodograms after randomly dropping 10 per cent of the data points each time. In the 90 per cent of the cases the highest peak is consistent with the 0.128\,d periodicity, which is likely related with the orbital period given the morphology of the light curve modulation. On the contrary, we do not see any significant peak in the region of periods suggested by \citet{zurita2015} (i.e. $3-6$\,cycle d$^{-1}$, note that we here assume an ellipsoidal variability with two maxima/minima per orbital cycle). Fitting a Gaussian model to the preferred peak in the periodogram yields $0.128 \pm 0.005$\,d, where the uncertainty corresponds to the standard deviation. In the case that the modulation is ellipsoidal (as it is in commonly observed in quiescent XRTs), the orbital period would be twice this value ($0.26 \pm 0.01$\,d = $6.24 \pm 0.24$\,h). In the top panel of Fig.~\ref{fig:phasefolded} we phase-folded the light curve with the $0.26$\,d period. A zero phase $T_0 (HJD) = 2458614.628 \pm 0.001$\,d was chosen so that the deepest minimum is placed at orbital phase 0.5, corresponding to the superior conjunction of the companion star. Two minima at different heights are present, supporting the potential ellipsoidal modulation. Additionally, the phase-folded light curve resulting from considering the 0.145-d peak shows significantly more scatter. Therefore, we hereafter adopt $0.26 \pm 0.01$\,d as the orbital period. However, we warn that an independent confirmation is necessary before accepting this as the definite orbital period of KY TrA.

\section{Discussion}
\label{sec:4}

\subsection{Constraints on the companion spectral type, the mass ratio and the distance of KY TrA}
\label{sec:spectraltype}
The spectral type of the donor star can be restricted by following the relation between the orbital period and the mean stellar density found for Roche lobe filling stars: $\bar{\rho} \approx 110 \times P^{-2}$\,gr cm$^{-3}$, where $P$ is the orbital period in hours \citep{frank2002}. Our orbital period ($0.26 \pm 0.01$\,d) implies a stellar density $2.06$\,gr cm$^{-3}$ which is consistent with a K5 V star \citep{drilling2000}. 

On the other hand, colour information of KY TrA can help to establish an upper limit on the spectral type, as we are disregarding any light contribution from a residual accretion disc. We took the photometry from five consecutive pairs of $r$ and $i$ images obtained over the two nights, and compute the colour $(r-i)$ of each set (see Table \ref{tab:colours}). This results in a weighted mean $(r-i) = 0.63 \pm 0.06$. The reddening of KY TrA is quite uncertain. \citet{murdin1977} quoted a lower limit $E(B-V)>0.5$ by comparison with the nearby Circinus field and observations of comparison stars near KY TrA. A rough estimate of the interstellar reddening towards KY TrA can be obtained from the IRAS and COBE/DIRBE dust maps\footnote{\url{https://irsa.ipac.caltech.edu/applications/DUST/}} \citep{schlegel1998}, re-calibrated with the Sloan DSS survey \citep{schlafly2011}. This yields $E(B-V) = 0.61 \pm 0.02$, in agreement with the lower limit proposed by \citet{murdin1977}. Correcting the interstellar extinction by applying $A_r = 2.285 \times E(B-V)$ and $A_i = 1.698 \times E(B-V)$ \citep{schlafly2011}, results in $(r-i)_0 = 0.27 \pm 0.08$ which is consistent with a $\approx$K2 V star \citep{covey2007}. This spectral type is slightly earlier than the K5 V star inferred from the orbital period-density relation, but this comes as no surprise given that we have so far neglected any contribution from the accretion disc to the quiescent light and also because donor stars are found to be undermassive for their spectral type \citep{kolb2001}. Our results are in line with the spectral type constraint K0-2 (or later) found by \citet{zurita2015} using $V-R$\,, $V-I$ colours.

The constraint on the spectral type of the companion star allows us to set a constraint on the mass ratio. By adopting a very conservative G8 V star classification, as implied by the upper limit on our de-reddened $(r-i)$ colour, we find $M_2 \leq 0.94$\,M$_\odot$ \citep{pecaut2013}\footnote{\url{https://www.pas.rochester.edu/~emamajek/EEM_dwarf_UBVIJHK_colors_Teff.txt}}. Since the compact object is expected to be a black hole, $M_1 \geq 3$\,M$_\odot$, then $q \leq 0.31$. 

Finally, from the orbital period and our quiescent magnitude we can provide a rough estimate of the distance to KY TrA. For this we employ the empirical correlation between $P$ and the absolute $r$-band magnitude $M_r$ given by equation 8 of \citet{casares2018}. This leads to $M_r = 6.8 \pm 0.2$, which can be related to the observed magnitude $r = 22.7 \pm 0.2$, interstellar extinction $A_r = 1.38$ and distance $d$ through the distance modulus equation. These numbers result in $d = 8.0 \pm 0.9$\,kpc. For comparison, \citet{vanparadijs1984} quote $d = 4.4$\,kpc\ based on the 1974 outburst, while \citet{murdin1977} propose $d > 3$\,kpc\ and $d \approx 7$\,kpc\ after drawing analogies with the outburst properties of A0620-00.

\begin{table}
	\centering
	\caption{Observed quiescent $(r-i)$ colours of KY TrA from the five consecutive pairs of $r$ and $i$ images. The mean HJD of each pair is also displayed.}
	\label{tab:colours}
	\begin{tabular}{ | c | c | } 
		\hline
            $(r-i)$ & mid-HJD \\ \hline 
		$0.57 \pm 0.13$ & 
	    \begin{tabular}{c} $2458614.56763$ \\
            \end{tabular} \\ \hline
            $0.76 \pm 0.10$ &
		\begin{tabular}{c} $2458614.66946$ \\
            \end{tabular} \\ \hline
            $0.48 \pm 0.13$ &
		\begin{tabular}{c} $2458614.82634$ \\
            \end{tabular} \\ \hline
		$0.83 \pm 0.15$ &
		\begin{tabular}{c} $2458615.65925$ \\
            \end{tabular} \\ \hline
            $0.42 \pm 0.14$ &
		\begin{tabular}{c} $2458615.84079$ \\
            \end{tabular} \\ \hline
	\end{tabular}
\end{table}

\subsection{The orbital inclination}
\label{sec:4.2}
The phase folded $r$-band light curve of KY TrA (top panel in Fig.~\ref{fig:phasefolded}) shows a maximum peak-to-peak amplitude of $\sim 0.5$\,mag, which is larger than the value expected for low-mass XRTs with an extreme 90 deg inclination \citep{bochkarev1979}. The light curve is also contaminated by flickering activity, as it is commonly the case in quiescent XRTs 
\citep[e.g.][]{zurita2003, yanes2022}. In addition, the amplitude seems  to change over different cycles as it appears flatter on our second observing night (Fig.~\ref{fig:LC}). We do not have a clear explanation for such complex behaviour although we speculate 
with the presence of a running superhump wave that may contaminate the ellipsoidal variability. This may enhance or dilute the amplitude of the ellipsoidal light curve depending on its relative phasing with respect to the orbital motion \citep{zurita2002}.

Despite these complexities, as an exercise, we have attempted to model the $r$-band light curve using the \textsc{xrbinary} code\footnote{Software developed by E.~L. Robinson, see \url{http://www.as.utexas.edu/~elr/Robinson/XRbinary.pdf} for further details.}. For this purpose, we adopted a Roche-lobe
filling companion with $T_{\rm{eff}} = 4440$\,K\ as the orbital period-density relation suggests a $\sim$K5 V companion (see Section \ref{sec:spectraltype}). In an effort to avoid the bias introduced by outlier data points (see top panel in Fig.\ref{fig:phasefolded}) we performed a $2 \sigma$ clipping routing: we first fit the ellipsoidal modulation, then remove the data points that deviate more than $2 \sigma$ from the fit and perform a new fit to the resulting curve. This procedure was iterated two times. The cleaned phase-folded light curve and the best-fitting model is shown in the middle panel of Fig.~\ref{fig:phasefolded}. The model gives an inclination of $74^{+8}_{-5}$\,deg, but with a fractional contribution of the donor star to the relative flux of 100 per cent which is larger than expected since no clear evidence for photospheric absorptions is seen in our spectroscopy. In addition, the equivalent width (EW) of the H$\alpha$ line is typically small in systems with a large contribution from the donor star because the disc light is diluted. For example, in GS 2000+25 where the donor star contribution is about 90 per cent, the EW of the H$\alpha$ line is $\mathbf{22 \pm 7}$\,\AA\ \citep{casares2015}, while we measure $66 \pm 6$\,\AA\ EW in KY TrA. Thus, we would expect the contribution of the companion to the $r$-band light curve to be $\lesssim 90$\,per cent in our system. We also modelled the phase-folded light curve binned with 0.03 phase bins and its best-fitting model is also presented in the bottom panel of Fig. \ref{fig:phasefolded}. For this fit we obtained an inclination of $78^{+11}_{-6}$\,deg, which is consistent with the result derived for the non binning light curve. In any case, given the large uncertainties, including possibly large systematics, we decide not to adopt the inclination constraints derived through the ellipsoidal modeling.

Alternatively, we tried to estimate the orbital inclination, $i$, through the empirical correlation found by \citet{casares2022} between $i$ and the depth of the trough, $T$, of the two peaks of the H$\alpha$ emission profile:
\begin{equation}
    i \,(\rm{deg}) = 93.5(6.5)\,T + 23.7(2.5)~, 
	\label{eq:inclination}
\end{equation}
with $T$ given by 
\begin{equation}
    T = 1 - 2 ^{1-\left(\frac{DP}{W}\right)^2}~,
	\label{eq:depth}
\end{equation}
where $W$ is the FWHM of the symmetric two-Gaussian model fit to the H$\alpha$ profile. From the symmetric double Gaussian fit reported in Section~\ref{sec:3.1} we obtain $W = 972 \pm 31$\,km s$^{-1}$. Following \citet{casares2022} we have applied a small +0.01 systematic shift to the $T$ value in order to correct for instrumental resolution degradation. We obtained $T = 0.36 \pm 0.05$. Note that the phase coverage of our spectroscopy is below 50\,per cent of the orbit, and this could introduce some bias in the estimation of $i$ from the correlation. In order to account for the lack of the orbital modulation in $T$, we have measured this variability in two XRTs from \citet[][see details in \textbf{their} Appendix B]{casares2022} whose orbital periods bracket that of KY TrA (i.e. A0620-00 and GRO J0422+32). We find that the orbital variability of $T$ has a mean rms of 0.13, which we take as the systematic error that we will add quadratically to our $T$ measurement in KY TrA. Therefore, we establish $T = 0.36 \pm 0.14$, which implies an orbital inclination of $i = 57 \pm 13$\,deg\ through equation \ref{eq:inclination}.

\subsection{Monte Carlo simulations on binary parameters and stellar masses}
In order to constrain the binary parameters and stellar masses in KY TrA we have run a Monte Carlo simulation using as priors all the information assembled in this paper. From the $r$-band light curve we find evidence for a likely orbital period of $0.26 \pm 0.01$\,d while the H$\alpha$ emission line profile provides $K_2 = 501 \pm 52$\,km s$^{-1}$. This yields a mass function $f(M_1) = 3.2 \pm 1.0$\,M$_\odot$ (68 per cent confidence level), which represents a lower limit to the mass of the compact star.  

On the other hand, the de-reddened $(r-i)$ quiescent colour supports a K2V (or later) donor star but we adopt instead a conservative upper limit to the spectral type G8, based on the most extreme possible colour. This implies $M_2 \leq 0.94$\,M$_\odot$ and $q \leq 0.31$. Note that this upper limit on $q$ is very conservative since XRTs with low-mass donor stars show a very narrow distribution of mass ratios centered at $q \simeq 0.06$ \citep{casares2015, Casares2016}. The largest mass ratio in a XRT with a low-mass companion is actually found in GX339-4, with $q = 0.18 \pm 0.05$ \citep{heida2017}. Moreover, a lower limit on $q$ of $0.01$ can be adopted considering that no XRT has been found with a smaller mass ratio \citep[see][]{Casares2016}. Thereby, we adopt a normal distribution for $q$ ranging between $0.01 - 0.31$. Additionally, we establish a binary inclination of $i = 57 \pm 13$\,deg using the correlation between this parameter and $T$ (equation~\ref{eq:inclination}). Employing all these constraints, the masses of the stellar components can now be obtained from: 
\begin{equation}
    M_1 = \frac{f(M_1)(1+q)^2}{\sin^3{i}}; \; \; \; \; \; M_2 = qM_1
    \label{eq:massfun}
\end{equation}
A Monte Carlo simulation with $10^5$ trials have been computed. Given the loose constraints on mass ratio we find that a large number of solutions lead to impossible values that would imply spectral types earlier than G8. Following \citet{casares2023}, we decided to run a new Monte Carlo simulation with the priors $M_2 \leq 0.94$\,M$_\odot$ and the upper limit to the inclination set by $\cos{i} \geq 0.49q^{2/3}\left[0.6q^{2/3}+ \ln{\left(1+q^{1/3}\right)}\right]^{-1}$. The latter is a geometrical constraint that reflects the non detection of X-ray eclipses during outburst \citep{kaluzienski1975, murdin1977}. This results in $M_1 = 5.8^{+3.0}_{-2.4}$$\,M_\odot$ and $M_2 = 0.5 \pm 0.3$\,M$_\odot$ with a $68$ per cent confidence level, supporting a black hole nature for the compact object. As an independent test, the black hole mass can also be estimated applying the relation found in \citet{casares2022} between this parameter, $W$ and the orbital period following:
\begin{equation}
    M_1^* = 3.45\times10^{-8} P_{\rm{orb}} \left(\frac{0.63W+145}{0.84}\right)^3   \rm{M_\odot}~,
	\label{eq:mass}
\end{equation}
with $P_{\rm{orb}}$ expressed in days. This yields $M_1^* = 6.6 \pm 0.6$\,M$_\odot$, which agrees well with the value obtained with equation~\ref{eq:massfun}.

\section{Conclusions}
\label{sec:5}
We present a new optical study of the black hole X-ray transient KY TrA based on time-resolved FORS2 spectroscopy, $0.5$\,arcsec resolution imaging and DECam photometry. We have obtained refined astrometric coordinates for KY TrA and the line-of-sight field star separated only $0.64 \pm 0.04$\,arcsec from the XRT. We derived the orbital parameter $K_2 = 501 \pm 52$\,km s$^{-1}$ by exploiting an empirical correlation with the FWHM of the H$\alpha$ emission line. The $r$-band light curve, on the other hand, presents variability consistent with an ellipsoidal modulation. By applying a Lomb-Scargle periodogram we obtain a likely orbital period of $0.26 \pm 0.01$\,d. These parameters imply a mass function $f(M_1) = 3.2 \pm 1.0$\,M$_\odot$. 

In addition, the de-reddened quiescent colour $(r-i)$ is consistent with a $\approx$K2 spectral type donor star (or later, in case of significant accretion disc contamination). By adopting the most extreme colour we find that the companion must have a spectral type later than G8 which translates into upper limits of $M_2 \leq 0.94$\,M$_\odot$ and $q \leq 0.31$. Furthermore, the correlation between the depth of the H$\alpha$ line trough and the binary inclination led to $i = 57 \pm 13 $\,deg. All these constraints together with the non-detection of X-ray eclipses during outburst yield $M_1 = 5.8^{+3.0}_{-2.4}$$\,M_\odot$ and $M_2 = 0.5 \pm 0.3$\,M$_\odot$\,for the masses of the compact and companion star, respectively. Our result confirms the presence of a black hole in the X-ray transient KY TrA, as it has been long suspected from the X-ray properties displayed during the 1974 and 1990 outbursts. In addition, we propose that KY TrA is located at $8.0 \pm 0.9$\,kpc. More future spectroscopic and photometric observations are required to better sample the orbit and derive more accurate constraints to the binary parameters, in particular to the mass ratio and orbital inclination.

\section*{Acknowledgements}
The {\sc molly} package developed by Tom Marsh is gratefully acknowledged. This work is supported by the Spanish Ministry of Science via an Europa Excelencia grant (EUR2021-122010) and the Plan de Generación de conocimiento: PID2020-120323GB-I00 and PID2021-124879NB-I00. 

\section*{Data Availability}
The FORS2 spectroscopy and DECam photometry data are available from \url{http://archive.eso.org/eso/eso_archive_main.html} and \url{https://astroarchive.noirlab.edu/portal/search/}, respectively.



\bibliographystyle{mnras}
\bibliography{references} 



\bsp	
\label{lastpage}
\end{document}